\documentclass[11pt,a4paper]{article}

\usepackage{fullpage}
\usepackage{indentfirst}
\usepackage{color}
\usepackage[utf8x]{inputenc}
\usepackage[T1]{fontenc}
\usepackage{hyperref}
\usepackage[english]{babel}
\usepackage{amsmath}
\usepackage{amsfonts}
\usepackage{amssymb}
\usepackage{amscd}
\def \be{\begin{equation}}
\def \ee{\end{equation}}
\def \nn{\nonumber}

\def \d{{\rm d}}
\DeclareMathOperator{\arccoth}{arccoth}
\newcommand{\parcial}[1]{ \frac{\partial}{\partial #1} }

\author{J. Guerrero$^{1,2}$, F.F. L\'opez-Ruiz$^{3,1}$, V. Aldaya$^{1}$ and F. Coss\'{\i}o$^{1}$}

\title{Symmetries of the quantum damped harmonic oscillator}

\date{\begin{center}
\begin{small}$^1$ Instituto de Astrof\'{\i}sica de Andaluc\'{\i}a, IAA-CSIC,
\end{small}\\
\begin{small}
  Apartado Postal 3004, 18080 Granada, Spain
\end{small}\\
\begin{small}$^2$Departamento de Matem\'atica Aplicada, Universidad de
Murcia, \end{small}\\
\begin{small}Campus de Espinardo, 30100 Murcia, Spain.\end{small}\\
\begin{small}$^3$Departamento de F\'\i sica Aplicada, Universidad de
C\'adiz, \end{small}\\
\begin{small}Campus de Puerto Real, 11510 Puerto Real, C\'adiz, Spain.\end{small}\\
\begin{small}juguerre@um.es\ flopez@iaa.es\ valdaya@iaa.es\ fcossiop@gmail.es
\end{small}\\
\end{center}
}
\begin{document}

\maketitle
%


\begin{abstract}

For the non-conservative Caldirola-Kanai system, describing a quantum damped harmonic oscillator, a
couple of constant-of-motion operators generating the Heisenberg-Weyl algebra can be
found. The inclusion of the standard time evolution generator (which is not a symmetry) as a symmetry in
this algebra, in a unitary manner, requires a
non-trivial extension of this basic algebra and hence of the physical system itself.
Surprisingly, this extension leads directly to the so-called Bateman dual system,
which now  includes a new particle acting as an energy reservoir. In addition, the
Caldirola-Kanai dissipative system can be retrieved by imposing constraints.
The algebra of symmetries of the dual system is presented, as well as a
quantization that implies, in particular, a first-order Schr\"odinger equation.
As opposed to other approaches, where it is claimed that the spectrum of the Bateman Hamiltonian is complex and discrete,
we obtain that it is real and continuous, with infinite degeneracy
in all regimes.

\end{abstract}

\section{Introduction}

The interest in dissipative systems at the quantum level has remained constant
since the early days of Quantum Mechanics. Within its framework, the attempts to
describe damping, which intuitively could be understood as a mesoscopic property, have motivated a huge amount of papers, 
both from a theoretical and a more applied points of view.

Applications of quantum dissipation include quantum optics  \cite{Scully-Zubairy} and the study of decoherence phenomena \cite{Breuer}. 
It is customary to model dissipation by means of the theory of open systems or the thermal bath approach, in which a damped system is 
considered to be a subsystem of a bigger one with many, or infinite, degrees of freedom \cite{Weiss}. However, damped systems are
interesting in themselves as fundamental ones if one wants to capture the essential dissipative behavior in a simple model. In particular, 
the quantum damped harmonic oscillator has attracted much attention, as could be considered one of the simplest and paradigmatic examples 
of non-conservative system.

The quantum damped harmonic oscillator has been described frequently by the Caldirola-Kanai model \cite{Caldirola,Kanai}, which includes 
a time-dependent Hamiltonian. It has been claimed that, in the quantum theory built out of this kind of models, uncertainty  relations are 
not preserved under time evolution and could eventually be violated  \cite{Brittin,Razavy_libro}, but this apparent inconsistency is 
associated with  a confusion between canonical momentum and ``physical'' momentum (see, for instance, \cite{Schuch}; in fact, the same 
author proposed a logarithmic nonlinear Schr\"odinger equation to deal with this situation \cite{Schuch2}). On the contrary, the main 
drawback of these models might be that they do not describe the interplay between the system and the environment, so that phenomena 
such as decoherence are not taken into account. It is nevertheless interesting to note that this kind of models can in fact be derived 
from a system-plus-reservoir model
\cite{Yu-Sun_PhysRevA1994,Sun-Yu_PhysRevA1995} in the case of Ohmic spectral density.

The analysis of damping from the symmetry point of view has proved to be
especially fruitful. In a purely classical context, the symmetries of the
equation of the damped harmonic oscillator with time-dependent parameters were
found in  \cite{Martini}. For the usual damped harmonic oscillator,
finite-dimensional point symmetry groups for the corresponding Lagrangian (the un-extended Schr\"odinger group \cite{Niederer1,Niederer2,Niederer3}) 
and the equations of motion ($SL(3,\mathbb{R})$) respectively, were found in \cite{Cervero-C,Cervero-Q},
as well as an infinite contact one for the set of trajectories of the classical equation.

As far as the quantum theory is concerned, in \cite{QAT} the authors provided a neat framework to study a
generalized version of the Caldirola-Kanai model with time-dependent parameters (frequency, damping coefficient
and external force), based on a quantum version of the Arnold transformation (QAT) \cite{Arnold}. The integrals
of motion and symmetries were identified and exploited to calculate wave functions, basic operators and the exact
time evolution operator, generalizing some of the results in \cite{Manko}. Many papers related to the Caldirola-Kanai
model keep appearing \cite{Um-Yeon,Fresneda}, showing that the debate about fundamental quantum damping is far from
being closed (the reader can see \cite{QAT} for further references on the Caldirola-Kanai model). For
instance, at the quantum level of this model, it remains to address the symmetry role of the time translation generator 
$i\hbar \parcial{t}$. In fact, $i\hbar \parcial{t}$ acting on a solution of the Caldirola-Kanai
Schr\"odinger equation is no longer a solution. As a consequence, the time evolution operator $\hat{U}$ does not
constitute a one-parameter group of unitary transformations.  Equivalently, the solution of the equation
$i\hbar\parcial{t}\hat{U}=\hat{H}(t)\hat{U}$ is not
$e^{-\frac{i}{\hbar}t \hat{H}(t)}$, but rather the time ordered product
$Te^{-\frac{i}{\hbar}\int\hat{H}(t)dt}$, referring to the Neumann series,  or
the Magnus series $e^{-\frac{i}{\hbar}\hat{\Omega}(t)}$ \cite{Magnus}.

The purpose of this article is to throw some light on the subject of quantum
dissipation with the guide of symmetry. The starting point will be the Caldirola-Kanai system
and we shall see that this model contains, somehow, a reminiscence of more general, system-plus-reservoir models.
Using the QAT, there can be found basic integrals of the motion closing a Heisenberg-Weyl algebra, imported from those of
the free particle, together with their squares. Time translations in the damped system do not belong to the imported
quadratic, conserved operators. This is to be expected, as the classical equation of motion includes a friction term
and the energy in this system is not conserved. The following question immediately arises: is there any
finite-dimensional group of symmetries containing time translations and, at least, the basic operators?
The answer is ``yes'', and we shall pay attention to this question in the case of the damped harmonic oscillator
and the surprising consequences of the subsequent calculation: for this symmetry to act properly, it is necessary
to enlarge the physical system with a new degree of freedom, corresponding to a new particle with interesting
properties. 

In his original paper \cite{Bateman}, Bateman looked for a variational
principle for equations of motion with a friction term linear in velocity, but
he allowed the presence of extra equations. This trick effectively doubles the
number of degrees of freedom, introducing a time-reversed version of the
original damped harmonic oscillator, which acts as an energy reservoir and could
be considered as an effective description of a thermal bath, as opposed to
the standard approach followed in the theory of open quantum systems, where the
thermal bath contains many, or infinite, degrees of freedom. The Hamiltonian
that describes this system was later rediscovered by Feschbach and Tikochinsky
\cite{Tikochinsky,Fesh-Tiko,Feschbach_libro,Dekker_report} and the corresponding
quantum theory was immediately analyzed.

Some issues regarding Bateman's dual system arose in the widespread
Feschbach-Tikochinsky construction of the corresponding Hilbert space and  spectrum. This construction leads to a set of complex 
eigenvalues of the Hamiltonian (see \cite{Chruscinski} and references therein), and it has been interpreted that the vacuum 
of the theory decays with time \cite{Celeghini}\footnote{We feel that the ultimate reason of this interpretation is the lack 
of a vacuum representation of the relevant symmetry algebra
(see Section~\ref{sec:arnold.newgroup}).}. In this paper we shall provide a rigorous construction of the physical Hilbert 
space and the real spectrum of the Hamiltonian, which are not found by means of the Feschbach-Tikochinsky construction. 
To do so, we shall take advantage of the
adoption of a canonical quantization in which the Schr\"odinger equation is first-order.
This approach is very similar to that presented in \cite{Rideau-etal} and it also has some relationship with the one in \cite{Chruscinski}; there, the generalized eigenvectors corresponding to the 
complex eigenvalues of the Hamiltonian are interpreted as resonant states.

Bateman's dual system is still discussed frequently
\cite{Blasone,Majima-Suzuki_AnnPhys2011}. Many authors
have considered this model as a \textit{good} starting point for
the formulation of the quantum theory of dissipation. One of the aims of this
paper will be to show that the study of the symmetries of the Caldirola-Kanai
model leads to Bateman's dual system, so that it should be considered as a
\textit{natural} starting point for the study of quantum dissipation. This could be seen as a ``step-back-and-forth'' 
approach (from a model of bath, then the Caldirola-Kanai model and finally Bateman's dual system), arriving at an idealization 
which intends to capture some of the features of quantum dissipation.
Moreover, we shall show that it is possible to get back to the Caldirola-Kanai system by imposing constraints in
the Bateman classical system.

The paper is organized as follows. We begin in Section~\ref{introarnold} by recalling the results from \cite{QAT},
in the case of the damped harmonic oscillator with constant coefficients in order to import basic operators
from the free particle system, which satisfy the condition of being integrals of the motion and close a
Heisenberg-Weyl algebra.
In Section~\ref{sec:arnold.newgroup} we look for a minimal algebra of operators which includes the basic conserved operators of the Caldirola-Kanai system and the generator of time translations $\frac{\partial\, }{\partial t}$.  In so doing, we arrive at an algebra of conserved operators of an enlarged physical system, which turns out to be the
Bateman dual system. We also show how, at the classical level, the Caldirola-Kanai system can be recovered 
from the Bateman system by a constraint.

In Section~\ref{sec: quantbat} we revise the quantization of the Bateman dual system. In
particular, we find a first-order Schr\"odinger
equation, from which the wave functions and the energy spectrum can be obtained.
Finally, an Appendix is devoted to the study of a non-minimal, infinite-dimensional symmetry algebra for the damped particle.

\section{Basic operators in the Caldirola-Kanai model}
\label{introarnold}

The classical equation of motion describing a Damped Harmonic Oscillator (DHO)
is given by:
\begin{equation}
\ddot{x}+\gamma\dot{x}+\omega^{2}x=0\,,
\label{eq:ecuacionDHO}
\end{equation}
where $\omega$ is a frequency and $\gamma$ is the damping constant,
both defining the system. This equation can be derived, in particular, from the Hamilton
principle with a time-dependent Lagrangian. The corresponding, time-dependent,
Hamiltonian function, to be named $H_{CK}$, is given by
\begin{equation}
  H_{CK} = e^{-\gamma t}\frac{p^2}{2m}+\frac{1}{2}m\omega^{2}x^{2} e^{\gamma
t}\,.
\end{equation}

The canonical quantization of $H_{CK}$ leads to the Caldirola-Kanai equation \cite{Caldirola,Kanai}, which is
a Schr\"odinger equation for the DHO:
\begin{equation}
i\hbar\frac{\partial \phi}{\partial t} = \hat H_{CK}\,\phi \equiv
-\frac{\hbar^{2}}{2m} e^{-\gamma t}\frac{\partial^{2} \phi}{\partial x^{2}}
+\frac{1}{2}m\omega^{2}x^{2} e^{\gamma t}\phi \,.
\label{eq:Schrodinger DHO}
\end{equation}
Some comments on the features of this equation have already been made in the
Introduction. Due to the explicit time dependence of the Hamiltonian function,
the corresponding operator $\hat H_{CK}$ is not a conserved operator and it is not possible to formulate an eigenvalue equation for it (that is, a
time-independent Schr\"odinger equation). Hence, there does
not exist a one-parameter group of time evolution. This is directly linked with the fact that the operator $\hat H_{CK}$ does 
not preserve the space of solutions of the equation (\ref{eq:Schrodinger DHO}) (see \cite{QAT} for a further discussion).

Now we are just interested in identifying basic position and
momentum operators associated with classical conserved quantities (Noether
invariants). In general, a conserved quantum operator $\hat O(t)$
must satisfy the relation:
\begin{equation}
  \frac{\d}{\d t} \hat O(t) \equiv \frac{\partial}{\partial t} \hat O(t)
   + \frac{i}{\hbar}[\hat H(t),\hat O(t)] = 0\,.
\label{eq:conservacion}
\end{equation}
These operators are also known as \textit{dynamical invariants}. Although the particular method to obtain these conserved operators is not
fundamental in what follows, we find especially useful for that purpose
the Quantum Arnold Transformation (QAT) technique developed in
\cite{QAT}.

The QAT for the DHO maps unitarily solutions of the  Schr\"odinger equation for
the Galilean free particle to solutions of (\ref{eq:Schrodinger DHO}) and, accordingly,
conserved operators in the Hilbert space of solutions
of the free Schr\"odinger equation to operators acting on the Hilbert space of
solutions of (\ref{eq:Schrodinger DHO}).

In particular, we are going to map the Galilean momentum operator $\hat \pi$,
corresponding to the classical conserved quantity `momentum', and the position
operator $\hat \kappa$, corresponding to the classical, conserved quantity
`initial position', to operators for the Caldirola-Kanai system. These
free operators are, explicitly:
\begin{equation}
\hat \pi = -i \hbar \frac{\partial}{\partial \kappa}\,,
\qquad \qquad
\hat \kappa =
\kappa + \frac{i \hbar}{m} \tau \frac{\partial}{\partial \kappa}\,,
\label{eq:operatas libres}
\end{equation}
that is, those basic, canonically commuting operators with constant expectation
values for the free evolution.

The only relevant part of the QAT in which we are interested here is how the above
operators are transformed (we refer the reader to \cite{QAT} for
further information). The QAT is defined just by two independent, classical
solutions of  (\ref{eq:ecuacionDHO}), $u_1(t)$ and $u_2(t)$, satisfying the
initial conditions  $u_1(0)=0,  u_2(0)=1, \dot{u}_1(0)=1, \dot{u}_2(0)=0$.
Their Wronskian $W(t) \equiv \dot{u}_{1}(t) u_{2}(t) -
u_{1}(t)\dot{u}_{2}(t)$ does not vanish. These classical solutions are:
\begin{equation}
u_{1}(t)=\frac{1}{\Omega}e^{-\frac{\gamma}{2}t}\sin\Omega t,\qquad
u_{2}(t)=e^{-\frac{\gamma}{2}t}\cos\Omega t +
\frac{\gamma}{2 \Omega}e^{-\frac{\gamma}{2}t}\sin\Omega t,
\label{eq:us DHO}
\end{equation}
for which  $W(t)= e^{-\gamma t}$, and
\begin{equation}
\Omega=\sqrt{\omega^{2}-\frac{\gamma^{2}}{4}}
\label{eq:frecuencia}
\end{equation}
is the characteristic frequency for the damped oscillator. Note that these solutions have good limit in the case of critical 
damping
$\omega = \tfrac{\gamma}{2}$ (i.e. $\Omega=0$).

Performing the QAT for the operators (\ref{eq:operatas libres}) results in the
formulas (see \cite{QAT}):
\begin{equation}
 \hat P =
 -i \hbar u_2 \frac{\partial}{\partial x} - m \frac{\dot u_2}{W} \, x\,,
\qquad
 \hat X =
  \frac{\dot u_1}{W}\, x +i \hbar \frac{u_1}{m}   \frac{\partial}{\partial x}\,.
\label{eq:operatas generales}
\end{equation}
It can be checked by direct computation that these expressions satisfy the
condition (\ref{eq:conservacion}) provided that $u_1$ and $u_2$ are solutions of
(\ref{eq:ecuacionDHO}). Explicitly:
\begin{align}
 \hat P &=
 -i \hbar \frac{e^{-\frac{\gamma t}{2}}}{2 \Omega }(2 \Omega \, \cos \Omega t
   +\gamma \, \sin \Omega t) \frac{\partial}{\partial x}
   + m \frac{e^{\frac{\gamma t}{2}} }{4 \Omega}
      \left(\gamma ^2+4 \Omega^2\right) \,\sin \Omega t \, x\,,
\label{eq:P_DHO}
 \\
 \hat X &=
  \frac{e^{\frac{\gamma t}{2}}}{2 \Omega } (2 \Omega \, \cos \Omega t
   -\gamma \, \sin \Omega t) \, x
    +i \hbar \frac{e^{-\frac{\gamma t}{2}}}{m \Omega }\, \sin \Omega t
     \frac{\partial}{\partial x}\,,
\label{eq:X_DHO}
\end{align}

with

\begin{equation}
  \left[ \hat X,\hat P \right] = i \hbar \,.
\end{equation}
They are the basic, canonically commuting operators with constant expectation
values (or, in other words, the two independent dynamical invariants) for the
evolution of the Caldirola-Kanai system describing the quantum DHO.

This way, we have imported the Heisenberg-Weyl algebra of conserved operators, which are symmetry generators, from the free 
particle system, but we have \textit{not} done so for a generator of time evolution. The time evolution generator present in 
the free particle, and which is a constant-of-motion operator, may also be brought into a constant-of-motion operator in the 
Caldirola-Kanai system 
and will be realized as a quadratic operator in $\hat P$ and $\hat X$ (more precisely, $\hat{P}^2$), but this does not represent 
the actual time evolution generator of the Caldirola-Kanai oscillator. It is desirable that the system be completely characterized by its 
symmetries, including its time evolution, but so far, this had not been the case.

\section{Addressing dissipative systems in a symmetry framework}
\label{sec:arnold.newgroup}

Even though it is possible to set up a clear framework to deal with the quantum
Caldirola-Kanai system for the DHO by employing the QAT, this does not provide by itself a well-defined
operator (on solutions) generating the actual time evolution. As mentioned previously, this is rooted in the 
fact that the conventional time evolution generator is
not included in the symmetry algebra:
the Hamiltonian does not preserve the Hilbert space of solutions of the Caldirola-Kanai
Schr\"odinger equation.

We would like to emphasize that the underlying motivation for constituting a closed algebra of quantum observables
is the general consensus that the quantization of any physical system must be a unitary and irreducible representation 
of some Poisson sub-algebra of classical observables characterizing the system to be quantized, irrespective of the 
particular method devoted to the achievement of this task (quantization method). The inclusion of the Hamiltonian in 
the chosen Poisson sub-algebra, and therefore in the quantum algebra, guarantees the characterization of the physical system.
One may accordingly wonder what happens if the generator of time evolution for the damped harmonic oscillator is forced 
to belong to the algebra containing the basic operators. We shall pursue this issue in this Section.

\subsection{Including time symmetry}
\label{timesymmetry}

In the Caldirola-Kanai model of the damped harmonic oscillator,
neither the operator $i\hbar\frac{\partial}{\partial t}$, nor
$\hat H_{CK}$ (which coincides with the former on solutions) 
preserve the space of solutions of the Schr\"odinger equation.
We shall impose
the condition of $i \hbar \frac{\partial}{\partial t}$ being a
symmetry generator. But it will be done in an elegant way, trying
to close with $\hat X$ and $\hat P$ an enlarged Lie algebra of (conserved, symmetry
generating) observables acting on the (possibly enlarged) Hilbert
space ${\cal H}$. Implicitly, we are forcing the system to
become conservative, but this is the only information we are going
to provide.

As a first step, we shall take advantage of the explicit
expressions obtained for the conserved basic operators: we compute
the commutators of $\hat{X}$ and $\hat{P}$ with $\hat{H} =
i\hbar\frac{\partial}{\partial t}$, and the resulting expressions
are considered as new operators. After that, the commutators of
the new operators with the formers are also computed, in the hope
that this process ends up closing a Lie algebra at a finite number
of steps. In fact, the resulting enlarged algebra is
finite-dimensional and includes $\hat{X}, \hat{P}, \hat{H}$ and
four more generators\footnote{At least in the simpler case of the damped
particle, infinitely many new generators can be included in its
Lie algebra. See Appendix~\ref{infinitegroup} for further
details.} (plus the central one $\hat{I}$), denoted by
$\hat{\tilde Q}, \hat{\Pi}, \hat{G}_1$ and $\hat{G}_2$,
\begin{align*}
 \hat P &=
 -i \hbar e^{-\frac{\gamma t}{2}} (\cos \Omega t
   +\frac{\gamma}{2 \Omega} \, \sin \Omega t) \frac{\partial}{\partial x}
   + m \,\frac{\omega^2}{\Omega} \,e^{\frac{\gamma t}{2}}
       \,\sin \Omega t \, x\,
 \\
 \hat X &=
  e^{\frac{\gamma t}{2}} (\cos \Omega t
   -\frac{\gamma}{2 \Omega} \, \sin \Omega t) \, x
    +i \hbar \frac{e^{-\frac{\gamma t}{2}}}{m \Omega }\, \sin \Omega t
     \frac{\partial}{\partial x}\,
 \\
 \hat \Pi &=
   - i \hbar e^{-\frac{\gamma t}{2}}\,
    (\cos \Omega t - \frac{\gamma}{2\Omega}\, \sin \Omega t)
    \frac{\partial}{\partial x}
   + m \frac{\omega^2}{\Omega } \,e^{\frac{\gamma t}{2}}\,
    \sin \Omega t \, x
 \\
 \hat{\tilde Q} &=
  e^{\frac{\gamma t}{2}} (\cos \Omega t
   -\frac{3 \gamma}{2 \Omega} \, \sin \Omega t) \, x
    +i \hbar \frac{e^{-\frac{\gamma t}{2}}}{m \Omega }\, \sin \Omega t
     \frac{\partial}{\partial x}\,
 \\
 \hat G_1 &=
   - \frac{1}{4 \Omega^2}\, (-4\omega^2 +\gamma^2 \cos 2 \Omega t
      + 2 \gamma \Omega \sin 2\Omega t) \,
\\
 \hat G_2 &=
    \frac{\gamma}{\Omega^2}\, \sin^2 \Omega t \,,
\end{align*}
which close the eight-dimensional algebra $\tilde{\mathcal A}$:
\begin{align}
  \left[\hat X, \hat P \right] &= i\hbar \hat{I}&
  \left[\hat{\tilde Q}, \hat \Pi \right] &= 2 i \hbar \hat G_1 - i
  \hbar\hat{I}\nonumber
\\
  \left[\hat X, \hat{\tilde Q} \right] &= -\frac{i\hbar}{m} \hat G_2&
  \left[\hat X, \hat \Pi \right] &= i\hbar \hat G_1 \nonumber
\\
  \left[\hat P, \hat{\tilde Q} \right] &= -i\hbar \hat G_1
                   + i\hbar \gamma \hat G_2&
  \left[\hat P, \hat \Pi \right] &= - i \hbar m \omega^2 \hat G_2 \nonumber
\\
  \left[\hat H, \hat X \right] &= -\frac{i\hbar}{m} \hat \Pi &
  \left[\hat H, \hat P \right] &= 2 i \hbar m \omega ^2 \hat X
                     - i \hbar m \omega^2 \hat{\tilde Q}
\label{algebra1}
\\
  \left[\hat H, \hat{\tilde Q} \right] &= -2 i\hbar \gamma \hat X
      -\frac{i\hbar}{m} \hat P + i\hbar \gamma \hat{\tilde Q} &
  \left[\hat H, \hat \Pi \right] &= 3 i\hbar m \omega^2 \hat X
     - 2 i\hbar m \omega^2 \hat{\tilde Q} - i\hbar \gamma \hat \Pi
     \nonumber
\\
  \left[\hat H, \hat G_1 \right] &= -i\hbar \gamma \hat G_1
                   + 2 i\hbar \omega^2 \hat G_2&
  \left[\hat H, \hat G_2 \right] &= - 2 i \hbar \hat G_1
                 + i \hbar \gamma \hat G_2 + 2 i \hbar \hat{I}\,. \nonumber
\end{align}

The outcome of this process depends, obviously, on the way
$\frac{\partial}{\partial t}$ acts on the explicit realization of
the Heisenberg-Weyl generators, which in this case encodes somehow
the time evolution of the Caldirola-Kanai oscillator ($\hat X$ and
$\hat P$ were conserved operators under the evolution of this
system).

Now we have to unveil the physical content of \eqref{algebra1}.
This algebra $\tilde{\mathcal A}$ corresponds to a central extension of an
algebra, to be named $\mathcal A$, and this particular central extension itself determines the actual basic
conjugated pairs, fixing a specific quantization. The
operators $\hat{\tilde Q}$ and $\hat{\Pi}$ (plus $\hat{I}$) expand
a Heisenberg-Weyl subalgebra, and $\hat{H}, \hat{G}_1$ and
$\hat{G}_2$ expand a 2-D affine algebra (with $\hat{H}$ acting as
dilations). However, in this realization $\hat{\tilde Q}$ and
$\hat{\Pi}$ are not basic operators\footnote{This can be seen as an anomaly, 
see \cite{Alda-Loll92,simplin}.}, while $\hat{H}$ and $\hat{G}_2$ are conjugate ones:
time proves to be a basic variable, this way contributing to 
the parametrization of the solution manifold. This result is
puzzling: time is not expected to be a coordinate or momentum of any degree of
freedom and therefore the corresponding generator should not
appear as an element of a conjugate pair. We then wonder whether it is possible or not to take
advantage of the information encoded in the algebra $\mathcal A$ to describe a
physical system in which we have an evolution with respect to the
ordinary time variable. It could not even be discarded the possibility that in the quantum 
reduction process, in which $\hat{\tilde Q}$ and $\hat{\Pi}$ cease to be basic operators, 
$\hat{H}$ would regain its expected role of time evolution generator. However, this requires a 
deeper analysis and fortunately this is not the only solution. 

In fact, $\tilde{\mathcal A}$ is not the only possible central extension of $\mathcal A$ and 
our strategy here will be to consider other possible quantizations of the un-extended
algebra $\mathcal A$, labeled by its possible central extensions.
A detailed study 
shows that there are three relevant kinds of central extensions,
describing systems with different degrees of freedom.
Thinking of the algebra above as an abstract Lie algebra, it can
be shown that a couple of parameters, in addition to the ordinary mass parameter $m$, control the central 
extensions that are allowed by the Jacobi identity of Lie algebras. We shall focus on a one-paremeter family $\tilde{\mathcal A}_k$
which contains the most relevant cases from the physical point of view:
\begin{align*}
  \left[\hat X, \hat P \right] &= i\hbar \hat{I}&
  \left[\hat{\tilde Q}, \hat \Pi \right] &= 2 i \hbar \hat G_1 - i \hbar
k \hat{I}
\\
  \left[\hat X, \hat{\tilde Q} \right] &= -\frac{i\hbar}{m} \hat G_2&
  \left[\hat X, \hat \Pi \right] &= i\hbar \hat G_1
\\
  \left[ \hat P,\hat{\tilde Q} \right] &= -i\hbar \hat G_1
                   + i\hbar \gamma \hat G_2 + i\hbar (k-1)\hat{I}&
  \left[\hat P, \hat \Pi \right] &= - i \hbar m \omega^2 \hat G_2
\\
  \left[\hat H, \hat X \right] &= -\frac{i\hbar}{m} \hat \Pi &
  \left[\hat H, \hat P \right] &= 2 i \hbar m \omega ^2 \hat X
                     - i \hbar m \omega^2 \hat{\tilde Q}
\\
  \left[\hat H, \hat{\tilde Q} \right] &= -2 i\hbar \gamma \hat X
      -\frac{i\hbar}{m} \hat P + i\hbar \gamma \hat{\tilde Q}&
  \left[\hat H, \hat \Pi \right] &= 3 i\hbar m \omega^2 \hat X
     - 2 i\hbar m \omega^2 \hat{\tilde Q} - i\hbar \gamma \hat \Pi
\\
  \left[\hat H, \hat G_1 \right] &= -i\hbar \gamma \hat G_1
                   + 2 i\hbar \omega^2 \hat G_2&
  \left[\hat H, \hat G_2 \right] &= - 2 i \hbar \hat G_1
                 + i \hbar \gamma \hat G_2 + i\hbar (1+k)
                 \hat{I}\,.
\end{align*}

It is now convenient to perform the linear shift:
\[
  \hat Q \equiv  \hat{\tilde Q} + (k-1) \hat X\,,
\]
so that the actual degrees of freedom diagonalize:
\begin{align*}
  \left[\hat X, \hat P \right] &= i\hbar \hat{I}&
  \left[\hat Q, \hat \Pi \right] &= i \hbar (k+1) \hat G_1 - i \hbar k \hat{I}
\\
  \left[\hat X, \hat Q \right] &= -\frac{i\hbar}{m} \hat G_2&
  \left[\hat X, \hat \Pi \right] &= i\hbar \hat G_1
\\
  \left[\hat P, \hat Q \right] &= -i\hbar \hat G_1
                   + i\hbar \gamma \hat G_2 &
  \left[\hat P, \hat \Pi \right] &= - i \hbar m \omega^2 \hat G_2
\\
  \left[\hat H, \hat X \right] &= -\frac{i\hbar}{m} \hat \Pi &
  \left[\hat H, \hat P \right] &= i \hbar m \omega ^2 (1+k) \hat X
                     - i \hbar m \omega^2 \hat Q
\\
  \left[\hat H, \hat Q \right] &=
     - i\hbar \gamma (1+k) \hat X -\frac{i\hbar}{m} \hat P &
  \left[\hat H, \hat \Pi \right] &=  i\hbar m \omega^2 (2k+1)\hat X
\\
      & \qquad + i\hbar \gamma \hat Q + \frac{i \hbar}{m}(1-k)\hat \Pi  \nn &
      & \qquad - 2 i\hbar m \omega^2 \hat Q - i\hbar \gamma \hat \Pi    \nn
\\
  \left[\hat H, \hat G_1 \right] &= -i\hbar \gamma \hat G_1
                   + 2 i\hbar \omega^2 \hat G_2  &
  \left[\hat H, \hat G_2 \right] &= - 2 i \hbar \hat G_1
                 + i \hbar \gamma \hat G_2 + i\hbar (1+k)
                 \hat{I}\,.
\end{align*}

By noticing the appearance of the central generator $\hat I$ on the right hand side of these commutation relations, 
we can see that the centrally extended algebras $\tilde{\mathcal A}_k$ can be classified as follows:
\begin{itemize}
\item  For $k \neq \pm 1$, a generic family $\tilde{\mathcal A}_k$ describes systems  with 3 degrees of freedom:
$(\hat{X},\hat{P})$,
$(\hat{Q},\hat{\Pi})$ and
$(\hat{H},\hat{G}_2)$, thus time being a conjugate variable.

\item For $k=1$, $\tilde{\mathcal A}_1=\tilde{\mathcal A}$ (already considered)
describes an anomalous system with 2 degrees of freedom:
$(\hat{X},\hat{P})$
and $(\hat{H},\hat{G}_2)$, time being again a conjugate variable.

\item For $k=-1$, $\tilde{\mathcal A}_{-1}$ describes a system with just 2 degrees of freedom:
$(\hat{X},\hat{P})$ and
$(\hat{Q},\hat{\Pi})$.

\end{itemize}

The third case is what we are looking for: although it contains two
degrees of freedom, time is \textit{not} a conjugate variable to any other. We should stress at this point that $\tilde{\cal A}_{-1}$ is in fact the unique central 
extension, inside the family $\tilde{\cal A}_k$, in which $\hat{H}$ is directly  an ordinary time evolution generator\footnote{This is still true for the complete family of central extensions of the algebra ${\cal A}$.
In addition, for $k=-1$ both degrees of freedom share the same central extension parameters.}. This is the reason why the whole process is unambiguously defined. 

A complete study of the entire family of central extensions of ${\cal A}$ and specially the original case $k=1$ and the corresponding aforementioned anomalous reduction of the quantum representation, which might eventually end up with 
a non-conserved Hamiltonian operator associated with $\hat H$, will be published elsewhere. 

Explicitly, for $\tilde{\mathcal A}_{-1}$ we have:
\begin{align*}
  \left[\hat{X}, \hat{P} \right] &= i\hbar\hat{I} &
  \left[\hat{Q}, \hat{\Pi} \right] &=  i \hbar \hat{I}\nn
\\
  \left[\hat{X}, \hat{Q} \right] &= -\frac{i\hbar}{m} \hat{G}_2&
  \left[\hat{X},  \hat{\Pi} \right] &= i\hbar \hat{G}_1 \nn
\\
  \left[\hat{P},  \hat{Q} \right] &= -i\hbar \hat{G}_1
                   + i\hbar \gamma \hat{G}_2 &
  \left[\hat{P},  \hat{\Pi} \right] &= - i \hbar m \omega^2 \hat{G}_2 \nn
\\
  \left[\hat{H}, \hat{X} \right] &= -\frac{i\hbar}{m} \hat{\Pi} &
  \left[\hat{H}, \hat{P} \right] &= -i \hbar m \omega^2  \hat{Q}  \nn
\\
  \left[\hat{H}, \hat{Q}\right] &= \frac{i\hbar}{m} (-\hat{P}+2\hat{\Pi})
       &
  \left[\hat{H}, \hat{\Pi} \right] &= -i\hbar m \omega^2 (\hat{X}+2 \hat{Q})  \nn
\\ & + i\hbar \gamma \hat{Q} & &
      - i\hbar \gamma  \hat{\Pi}  \nn
\\
  \left[\hat{H}, \hat{G}_1 \right] &= -i\hbar \gamma \hat{G}_1
                   + 2 i\hbar \omega^2 \hat{G}_2&
  \left[\hat{H}, \hat{G}_2 \right] &= - 2 i \hbar \hat{G}_1
                 + i \hbar \gamma \hat{G}_2  \,.
\end{align*}

In this case we notice that the operators $\hat{G}_1$ and $\hat{G}_2$ commute
with the basic couples $(\hat{X},\hat{P})$ and $(\hat{Q},\hat{\Pi})$ and, therefore,
will be represented as constants (times the identity operator) in any irreducible representation. Even more, their commutation 
rules with $\hat H$ determine that the value of these constants must be zero
($[\hat{H}, \hat{G}_1 ]=[\hat{H}, \hat{G}_2 ] = 0$, as
$\hat{G}_1$ and $\hat{G}_2$ are proportional to the identity $\hat{I}$; then, solve for $\hat{G}_1$ and $\hat{G}_2$ using the 
expressions above\footnote{To be precise, in the critical-damping case $\Omega= \sqrt{\omega^2 - \frac{\gamma^2}{4}}=0$ this
only implies $\hat G_1=\frac{\gamma}{2}\hat G_2$, although they still can be  represented trivially.}). Technically, the operators 
$\hat{G}_1$ and $\hat{G}_2$ are \textit{gauge}, in the sense that, in the resulting physical system, their symmetry transformation
does not produce any change in the parameters associated with basic operators.
In consequence, the effective dimension of the algebra $\tilde{\mathcal A}_{-1}$ is
$5+1$. Let us denote by $\tilde{\mathcal B}$ the effective reduced Lie algebra, whose homonymous generators 
$(\hat{X},\hat{P})$, $(\hat{Q},\hat{\Pi})$, $\hat{H}$ and $\hat{I}$ have the commutation relations:
\begin{align}
  \left[\hat{X}, \hat{P} \right] &= i\hbar \hat{I} &
  \left[\hat{Q}, \hat{\Pi} \right] &=  i \hbar\hat{I} \nn
\\
  \left[\hat{X}, \hat{Q} \right] &= 0&
  \left[\hat{X},  \hat{\Pi} \right] &= 0\nn
\\
  \left[\hat{Q},  \hat{P} \right] &= 0 &
  \left[\hat{P},  \hat{\Pi} \right] &= 0 \nn
\\
  \left[\hat{H}, \hat{X} \right] &= -\frac{i\hbar}{m} \hat{\Pi} &
  \left[\hat{H}, \hat{P} \right] &= - i \hbar m \omega^2  \hat{Q}
\\
 \left[\hat{H}, \hat{Q}\right] &= \frac{i\hbar}{m} (-\hat{P}+2\hat{\Pi})
       &
  \left[\hat{H}, \hat{\Pi} \right] &= -i\hbar m \omega^2 (\hat{X}+2 \hat{Q})  \nn
\\ & \,\,\,\,\, + i\hbar \gamma \hat{Q} & &
     \,\,\,\,\, - i\hbar \gamma  \hat{\Pi} \nn \,.
\end{align}

This way, we end up with two pairs of independent, canonical
operators along with a Hamiltonian. Since we have selected the
representation in which $\hat{H}$ is not a basic operator, it can
be written in terms of the basic ones in an irreducible
representation (note that symmetrization has been imposed to solve
the ordering ambiguity, so as to achieve unitarity):
\begin{equation}
  \hat{H} = \frac{1}{m} \hat{\Pi} \hat{P}-\frac{\gamma}{2}( \hat{Q} \hat{\Pi}
+ \hat{\Pi} \hat{Q}) - \frac{\hat{\Pi}^2}{m}- m
\omega^2 \hat{X} \hat{Q}
              - m \omega^2 \hat{Q}^2\,.
\label{hnuestro}
\end{equation}

The physical content of this Hamiltonian can be analyzed classically
by solving the Hamilton equations of motion. After that, a canonical quantization 
would also be possible. However, it is interesting to establish the relationship of 
the physical system at which we have arrived with a known one: the new system, with 
two degrees of freedom, is actually the Bateman dual system.

\subsection{Bateman's dual system}

Let us perform the following linear transformation of operators:

\begin{align}
\hat X &= \hat y + \frac{1}{m \gamma} \hat p_y +\frac{1}{2} \hat x \nn\\
\hat P&= \hat p_x- m \frac{\gamma}{2} \hat y  -  m \frac{\omega^2}{\gamma} \hat x \nn \\
\hat Q &=  -\hat y - \frac{1}{m \gamma} \hat p_y +\frac{1}{2} \hat x\nn \\
\hat \Pi &=  \hat p_x+ m \frac{\gamma}{2} \hat y - m \frac{\omega^2}{\gamma} \hat x\,,
\end{align}
the inverse of which is given by:
\begin{align}
\hat x &= \hat X + \hat Q \nn\\
\hat p_x &=  \frac{1}{2} (\hat P +\hat \Pi) + m \frac{\omega^2}{\gamma} (\hat X + \hat Q) \nn \\
\hat y &=  - \frac{1}{m \gamma} (\hat P - \hat \Pi) \nn \\
\hat p_y &= \hat P - \hat \Pi +\frac{1}{2}m \gamma (\hat X - \hat Q) \,.
\end{align}
This transforms the Hamiltonian \eqref{hnuestro} into the so-called Bateman's dual Hamiltonian \cite{Bateman}
\be
\hat{H}=\frac{ \hat{p}_x   \hat{p}_y}{m}+\frac{\gamma}{2}( \hat{y} \hat{p}_y
-\hat{x} \hat{p}_x)+m \Omega
^2 \hat{x}\hat{y} \,.
\label{hbateman}
\ee
$\hat{H}$ closes a 5+1 dimensional
algebra with $(\hat{x},\hat{p}_x)$ and
$(\hat{y},\hat{p}_y)$:
\begin{align}
  \left[\hat{x}, \hat{p}_x \right] &= i\hbar \hat{I} &
  \left[\hat{y}, \hat{p}_y \right] &=  i \hbar\hat{I} \nn
\\
  \left[\hat{x}, \hat{y} \right] &= 0&
  \left[\hat{x},  \hat{p}_y \right] &= 0 \nn
\\
  \left[\hat{y},  \hat{p}_x \right] &= 0 &
  \left[\hat{p}_x,  \hat{p}_y \right] &= 0 \nn
\\
  \left[\hat{H}, \hat{x} \right] &= \frac{i\hbar}{m}(- \hat{p}_y
+m \frac{\gamma}{2}\hat{x})&
  \left[\hat{H}, \hat{p}_x \right] &= i \hbar( -\frac{\gamma}{2} \hat{p}_x+ m
\Omega^2 \hat{y} ) \nn
\\
 \left[\hat{H}, \hat{y}\right] &= \frac{i\hbar}{m}(- \hat{p}_x
-m \frac{\gamma}{2}\hat{y})
       &
  \left[\hat{H}, \hat{p}_y \right] &= i \hbar( \frac{\gamma}{2} \hat{p}_y+ m
\Omega^2 \hat{x} ) \,.
\label{batealgebra}
\end{align}

It can be checked that the corresponding classical Hamiltonian $H$ describes a damped particle $(x,p_x)$ and its
time reversal $(y,p_y)$ by  computing the second order classical equations out of the Hamilton equations:
\be
\ddot{x}+\gamma\dot{x}+\omega^{2}x=0\,,\qquad
\ddot{y}-\gamma\dot{y}+\omega^{2}y=0\,.\label{dualeq}
\ee

The system is conservative, so that our objective of including the generator of time
evolution among the symmetries has been accomplished, though at the cost of including a
new degree of freedom.

\subsection{Recovering Caldirola-Kanai system from the conservative Bateman system}
\label{back2caldi}

Before addressing the quantum description of Bateman's dual system in the following section, 
let us insist on the classical theory in this subsection.
It is interesting to realize how the Bateman dual system actually
generalizes the Caldirola-Kanai damped harmonic oscillator in the sense that a Caldirola-Kanai particle and its mirror image can be recovered by a canonical transformation.

In Bateman's original paper \cite{Bateman}, the Caldirola-Kanai Lagrangian was obtained by imposing the simple
constraint $y=e^{\gamma t}x$ (Dekker \cite{Dekker_report} also described an analogous process by using a 
complex canonical tranformation). Here we provide a (real) canonical transformation which takes the Bateman Hamiltonian to the difference of a couple of dual Caldirola-Kanai Hamiltonians. That is:    
\begin{align}
y'&=\frac{1}{\sqrt{2}} \left(y- e^{\gamma t}x + \frac{\gamma}{2m\Omega^2}p_x \right)\nn \\
p'_y&=\frac{1}{\sqrt{2}} \left( p_y - \frac{\omega^2}{\Omega^2} e^{-\gamma t}p_x + m\frac{\gamma}{2}x \right)\nn \\
x'&=\frac{1}{\sqrt{2}} \left(x +
   e^{-\gamma t} y - \frac{\gamma}{2 m \Omega^2} e^{-\gamma t} p_x\right)\label{canonical-Bat2Cal}\\
p'_x &=\frac{1}{\sqrt{2}} \left(\frac{\omega^2}{\Omega^2} p_x + e^{\gamma t} p_y -
   m \frac{\gamma}{2} e^{\gamma t} x \right) \nn \\
t' &= t \,.\nn 
\end{align}
This transformation, although explicitly time dependent, is the lifting of a canonical transformation among initial constants. After performing the transformation, the new Hamiltonian
$H'$ (including the time derivative of the corresponding generating function) becomes:
\begin{equation}
 H' =  e^{-\gamma t'}\frac{p'_x{}^2}{2m}+\frac{1}{2}m\omega^{2}x'{}^{2} e^{\gamma t'} -
e^{\gamma t'}\frac{p'_y{}^2}{2m}-\frac{1}{2}m\omega^{2}y'{}^{2} e^{-\gamma t'} \,,
\end{equation}
which is nothing other than the Caldirola-Kanai Hamiltonian in terms of the variables $(x',p'_x)$ minus a dual Caldirola-Kanai Hamiltonian in terms of the variables $(y',p'_y)$. The relative minus sign in the new Hamiltonian $H'$ is unavoidable at least with a real canonical transformation. 


A constraint to obtain the original Caldirola-Kanai Hamiltonian now reduces simply to the form:
\begin{equation}
 y'=0\,, \quad p'_y = 0 \,.
\label{ligaduraprima}
\end{equation}


\section{Quantization of Bateman's dual system}
\label{sec: quantbat}

In principle, the quantum theory of the Bateman system can be addressed performing the usual canonical quantization in the position representation. Then, 
the Schr\"odinger equation for the Bateman Hamiltonian \eqref{hbateman} is given by: 
\be
i\hbar\frac{\partial\phi(x,y,t)}{\partial t}
=\left(-\frac{\hbar^2}{m}\frac{\partial^2}{\partial x\partial y}
-i\hbar \frac{\gamma}{2}(y\parcial{y} - x\parcial{x})+m \Omega ^2 x
y\right)\phi(x,y,t)\,.
\ee

It has been argued that the quantum Bateman system possesses
inconsistencies, such as complex eigenvalues and non-normalizable eigenstates 
\cite{Tikochinsky,Fesh-Tiko,Feschbach_libro,Dekker_report}. However, it should be noted that this 
observation is meaningless if the proper Hilbert space of the system is not specified, together with the precise way in which 
it is constructed. In this respect, Chru\'sci\'nski \& Jurkowski \cite{Chruscinski} showed that $\hat{H}$ has in fact a real,
continuous spectrum (we shall provide a proof of this in
Subsection~\ref{solving_equation}), and that the complex eigenvalues are
associated with
resonances, which in last instance are the responsible of dissipation.

\subsection{First-order Schr\"odinger equation}

In order to construct the quantum representation of the Bateman algebra \eqref{batealgebra}, we shall choose a mixed representation
of position-momentum on complex functions of $x$ and $p_y$: 
\begin{equation}
\hat x = x\,,\quad \hat p_x = - i \hbar \frac{\partial}{\partial x}\,,
\qquad \qquad
\hat y = i \hbar \frac{\partial}{\partial p_y} \,, \quad \hat p_y = p_y \,.
\label{operatas}
\end{equation}
With this choice to represent the basic operators, the Hamiltonian becomes: 

\begin{equation}
\hat H = i \hbar (\frac{\gamma}{2} x - \frac{p_y}{m}) \frac{\partial}{\partial x}
+ i \hbar (\frac{\gamma}{2} p_y + m \Omega^2 x) \frac{\partial}{\partial p_y}
+ i \hbar \frac{\gamma}{2}\,,
\label{qhbateman}
\end{equation}
that is, a first-order operator which is Hermitian with this particular ordering. Operators \eqref{operatas} and \eqref{qhbateman}
(together with the identity) are Hermitian and represent the algebra
\eqref{batealgebra} on complex, square integrable functions  
$\phi(x,p_y)$ with integration measure $\d x \d p_y$. 
As a consequence, the time-dependent Schr\"odinger equation becomes also first-order:
\begin{equation}
  i \hbar \frac{\partial \phi}{\partial t} = \hat H \phi =
  i \hbar (\frac{\gamma}{2} x - \frac{p_y}{m}) \frac{\partial \phi}{\partial x}
+ i \hbar (\frac{\gamma}{2} p_y + m \Omega^2 x) \frac{\partial \phi}{\partial p_y}
+ i \hbar \frac{\gamma}{2} \phi
\,.
\end{equation}

\subsection{Solving the equation: eigenfunctions and spectrum}
\label{solving_equation}

Before proceeding with the solution of the time-independent Schr\"odinger equation, let us firstly note that, being the 
Schr\"odinger equation first-order, it is possible to find the general form of the solution: 
\[
\phi(x,p_y,t) = e^{\frac{\gamma}{2}t} 
f\left(e^{\frac{\gamma}{2}t}(x \cos(\Omega t)-\frac{p_y}{m \Omega}\sin(\Omega t)),
e^{\frac{\gamma}{2}t}(p_y \cos(\Omega t)+ m \Omega x \sin(\Omega t)) \right)\,,
\]
where $f$ is an arbitrary complex function of two variables. This expression can be written in a more convenient 
form in order to compare with the solutions of the time-independent equation, by defining
\begin{equation}
z_{+}\equiv x + i \frac{1}{m \Omega} p_y \,,\qquad 
z_{-}\equiv x - i \frac{1}{m \Omega} p_y\,,
\label{zetas}
\end{equation}
which are complex and $z_{-}=z_{+}^*$ when $\Omega \in \mathbb R$ (the underdamping case).
Now, the solution is written:
\begin{equation}
\phi(x,p_y,t) = e^{\frac{\gamma}{2}t} 
f\left(e^{(\frac{\gamma}{2}+i \Omega)t}z_+,\;
e^{(\frac{\gamma}{2}-i\Omega t)} z_- \right)\,,
\label{solucionzetas}
\end{equation}
or even: 
\begin{equation}
\phi(x,p_y,t) = e^{\frac{\gamma}{2}t} 
f\left(e^{2 i \Omega t}\frac{z_-}{z_+},\;
e^{\gamma t} z_+ z_- \right)\,.
\label{solucionzetas2}
\end{equation}

Since this system is autonomous, it makes sense to formulate an eigenvalue problem for the Hamiltonian. 
It was already mentioned in the Introduction that several attempts of constructing the eigenfunctions 
of the Bateman dual system can be found in the literature. The most widespread is Feschbach-Tikochinsky's \cite{Tikochinsky,Fesh-Tiko}; they in fact looked for creation and annihilation operators to find the 
Hamiltonian eigenfunctions. This approach is appealing, as it unveils an $SU(1,1)$ structure 
for the system \cite{Celeghini}, although the fact that the Hamiltonian is not positive-definite leads to undesirable 
consequences: an unphysical complex, discrete spectrum along with the corresponding (non-normalizable, 
not even to a delta function) wave functions. 
It should be mentioned that many authors have nevertheless considered that this construction leads to the correct physical description of the quantum Bateman system \cite{Dekker_report,Celeghini}. 

Two sets of authors proposed independently a different approach to find the real part of the spectrum 
(and therefore,  the physically meaningful one) in the underdamping case. On the one hand, Rideau, Anderson 
and Hebda \cite{Rideau-etal} performed a canonical quantization in a mixed position-momentum representation (compare \eqref{solucionzetas} with their resutlts). On the other hand, 
Chrusci\' nski and Jurkowski \cite{Chruscinski} made use of angular variables to perform a first-order quantization 
of the  Hamiltonian, arriving at the correct spectrum, although it is not clear from their construction 
to what extent they can give a consistent construction of the basic operators (compare \eqref{solucionzetas2} with their resutlts).  We shall follow the first 
approach, although we shall give the corresponding expressions for every regime: underdamping, overdamping 
and critical damping. 

The general solution of the time-independent Schr\"odinger equation can be found in terms of the variables 
\eqref{zetas}. The eigenfunction of $\hat{H}$ with eigenvalue  $E$ is given in terms of an arbitrary 
function $g$ of a single argument:
\[
  \phi(z_{+},z_{-}) = \Bigl(\frac{z_{-}}{z_{+}}\Bigr)^{\frac{E-i\hbar \frac{\gamma}{2}}{2 \hbar \Omega}}
            g\bigl( z_{+} z_{-} \Bigl( \frac{z_{-}}{z_{+}}\Bigr)^{-\frac{i \gamma}{2 \Omega}}	
\bigr)\,.
\]
This means that the spectrum is infinitely degenerate. To break the degeneration, we proceed as usual: we select another operator 
which commutes with the Hamiltonian and find their simultaneous eigenfunctions. In fact, it is possible to split the Hamiltonian 
into two operators of that kind: 
\begin{align}
\hat H_\Omega &= -i \hbar \Big(\frac{p_y}{m} \frac{\partial}{\partial x}
- m \Omega^2 x \frac{\partial}{\partial p_y}\Big)\,, \label{opeHOme}
\\
\hat D &= i \hbar \frac{\gamma}{2}\Big( x  \frac{\partial}{\partial x}
+  p_y \frac{\partial}{\partial p_y} + 1\Big) \label{opeD}
\,, 
\end{align}
with:
\[
\hat H = \hat H_\Omega + \hat D\,,\qquad [\hat H_\Omega,\hat H] = 0\,,\qquad [\hat D,\hat H] = 0
\,.
\]
Then, the simultaneous eigenfunctions are, simply: 
\begin{equation}
\phi_{n,\lambda}(z_{+},z_{-}) = \Bigl(\frac{z_{-}}{z_{+}}\Bigr)^{\frac{n}{2}}
\Bigl( z_{+} z_{-} \Bigr)^{-\frac{1}{2}-i\lambda}\,,
\label{eigenfunctions}
\end{equation}
which satisfy:
\[
\hat H_\Omega \,\phi_{n,\lambda} = n \hbar \Omega \,\phi_{n,\lambda}\,, \qquad
\hat D \, \phi_{n,\lambda} = \lambda \hbar \gamma \, \phi_{n,\lambda} \quad
\Rightarrow \quad 
\hat H \, \phi_{n,\lambda} = \hbar (n  \Omega + \lambda  \gamma)\, \phi_{n,\lambda}\,,
\]
where the allowed values of $n$ and $\lambda$ depend on the regime.

\paragraph{Underdamping.}

In this case, $\Omega$ is real; therefore, $n$ must be real since the spectrum of $\hat H_\Omega$ must be real. Furthermore, 
$\frac{z_{-}}{z_{+}}$ is a pure phase, twice the one of $z_{-}$. So, $n$ is necessarily an integer. This may
 be seen as a consequence of the fact that 
$\hat H_\Omega$ represents a rotation in the $x,\,p_y$ plane for $\Omega^2 > 0$.

Regarding the value of $\lambda$, it must be real, given that $\gamma$ is real. No further constraints are 
required for  $\phi_{n,\lambda}$ to be well-defined. This is consistent with $\hat D$ representing a dilation, 
with spectrum the real line. So, the  spectrum of the Hamiltonian is:
\begin{equation}
 E = n \hbar \Omega + \lambda \hbar \gamma\,. 
\end{equation}
That is, we obtain a spectrum which is continuous and countably, infinitely degenerated.
These results coincide with those in \cite{Chruscinski}, although  they are
obtained here in a neater way: in \cite{Chruscinski} the authors quantize angular
variables and hence the basic operator ``multiply by the angle'' is not
defined. We have avoided this problem: our basic operators are just \eqref{operatas}.

\paragraph{Overdamping.}
Now $\Omega$ is pure imaginary, c.f. \eqref{eq:frecuencia}. In order to keep the spectrum real, $n$ must be pure imaginary as well. Let us then define
\[
 \tilde n \equiv -i n\,,\qquad \tilde \Omega \equiv -i \Omega\,.
\]
The variables \eqref{zetas} are now real:
\begin{equation}
z_{+}\equiv x + \frac{1}{m \tilde\Omega} p_y \,,\qquad 
z_{-}\equiv x - \frac{1}{m \tilde\Omega} p_y\,,  
\end{equation}
and it can be checked that solutions \eqref{eigenfunctions} can be written in the form:
\begin{equation}
 \phi_{n,\lambda}(z_{+},z_{-}) = e^{-i\tilde n \arccoth( \frac{m \tilde \Omega x}{p_y})}
\Bigl( z_{+} z_{-} \Bigr)^{-\frac{1}{2}-i\lambda}\,.
\end{equation}
Given that $\arccoth( \frac{m \tilde \Omega x}{p_y})$ is not a periodic function, $\tilde n$ can take any real value. This is consistent with the fact that the action of $\hat H_\Omega$ is analogous to that of a space-time boost in this regime.  

For the value of $\lambda$, the same considerations as before lead to the conclusion that it must be real. Therefore, the spectrum of the Hamiltonian is now real with two real indices: 
\begin{equation}
E = \tilde n \hbar \tilde \Omega + \lambda \hbar \gamma\,. 
\end{equation}
In other words, the spectrum in this case is continuous and uncountably, 
infinitely degenerated.

\paragraph{Critical damping.}

In order to find the eigenfunctions and the spectrum in the critical damping case, it is necessary to solve the eigensystem for the operators \eqref{opeHOme} 
and \eqref{opeD} when $\Omega$ is zero. This leads to eigenfunctions of the form: 
\begin{equation}
 \phi^o_{k,\lambda}(x,p_y) = e^{i k \frac{m \gamma x}{p_y}} (p_y^2)^{-\frac{1}{2}-i\lambda} \,, 
\end{equation}
satisfying ($\hat{H}_0=\hat{H}_{\Omega=0}$): 
\[
\hat H_0 \,\phi^o_{k,\lambda} =  k \hbar \gamma \,\phi^o_{k,\lambda}\,, \qquad
\hat D \, \phi^o_{k,\lambda} = \lambda \hbar \gamma \, \phi^o_{k,\lambda} \quad
\Rightarrow \quad 
\hat H \, \phi^k_{k,\lambda} = \hbar \gamma (k  + \lambda )\, \phi^o_{k,\lambda}\,,
\]
where $k$ and $\lambda$ must be real, so that the spectrum
\begin{equation}
E = \hbar \gamma (k + \lambda ) 
\end{equation}
is again continuous and uncountably, infinitely degenerated.

\bigskip

It is now clear that this quantization, which renders the Hamiltonian first-order, allows a straightforward construction of the Hilbert space of the quantum Bateman dual system as well as the corresponding physical spectrum, which is not complex but real, as it must be for a Hermitian Hamiltonian.

\section{Conclusions and Outlook}
\label{sec:conclusions}

In this paper we have found, starting with the non-conservative Caldirola-Kanai model for
the quantum damped harmonic oscillator, a larger system containing two 
degrees of freedom (the original system and its time reversal one) by imposing that the time evolution generator $i\hbar\frac{\partial\, }{\partial t}$ 
closes a symmetry algebra with the basic operators $\hat{X}$ and $\hat{P}$. The system obtained is the Bateman dual system which, in some sense, 
can be  considered as the simplest model for a system-reservoir approach to the damped harmonic oscillator.

It should be stressed that the process followed is unambiguous, in the sense that the symmetry algebra of the Bateman dual system is the only 
extension of the Caldirola-Kanai symmetry algebra where the time generator is not associated with a canonical variable. It would be worth investigating, however, the other possible
extensions of the Caldirola-Kanai symmetry algebra where this is not primarily the case, for both the two and the three degrees of freedom cases (see Section
\ref{timesymmetry}).

As commented in the Introduction, the quantization of Bateman's dual system has usually  been considered full of inconsistencies since  
Feschbach and Tikochinsky \cite{Tikochinsky,Fesh-Tiko,Feschbach_libro,Dekker_report} constructed its Hilbert space made of non-normalizable energy eigenstates with
complex discrete eigenvalues.
Many other authors insisted on this construction and suggested alternatives \cite{Celeghini,Blasone}, but in this paper we have shown that,
using a first-order Schr\"odinger equation, the quantization is well defined and the spectrum is real and continuous, with infinite degeneracy \cite{Rideau-etal,Chruscinski}.

The only pathology of the Bateman dual system is that the spectrum of the Hamiltonian is not bounded from below, but this is not a problem as long as the whole Bateman
system is isolated (i.e. there is no external interaction). This problem can be easily solved by restricting the initial conditions for the two
particles in such a way that the kinetic energy be positive \cite{Bateman}. This effectively describes a system with just one degree of freedom, 
as shown in Section \ref{back2caldi} by imposing the constraint \eqref{ligaduraprima} after
performing the canonical transformation \eqref{canonical-Bat2Cal}, which is equivalent to
a relation among initial constants.
It would be interesting to investigate 
the resultant one-degree-of-freedom systems after imposing
other relations among initial constants that give rise to positive kinetic energy.

The Bateman dual system solves many of the problems that the Caldirola-Kanai model for damped harmonic oscillator suffers from. 
For instance, while the time evolution under the Liouville equation for the density matrix of the Caldirola-Kanai system does not give rise to decoherence 
(one of the main points raised against the
Caldirola-Kanai model), the Bateman dual system suffers from decoherence and classical correlations after the mirror degree of freedom is traced out
\cite{Kim-etal}.
Also,  selecting appropriate dense subspaces of the Bateman Hilbert space, an effective evolution 
describing damped oscillations with complex discrete frequencies is obtained \cite{Rideau-etal}, and these complex discrete frequencies, appearing as
the eigenvalues of the Bateman dual Hamiltonian in the Feschbach-Tikochinsky quantization, turn out to be  resonant energies corresponding to
resonant (Gamow) states for the Bateman dual system \cite{Chruscinski}. Even more, an splitting of the Hilbert space can be made in such a way that 
in each subspace the originally unitary time evolution falls down to a semigroup, and therefore irreversible to an evolution \cite{Chruscinski}.
In addition, the Caldirola-Kanai system can be derived from Bateman's dual system  
 (see Section \ref{back2caldi}).

To end this section, we shall mention that the result that  the algebra $\tilde{A}$  \eqref{algebra1} is finite-dimensional depends
crucially on the fact that the frequency $\omega$ and the damping coefficient $\gamma$ are constant. For time dependent frequency 
$\omega(t)$ and damping coefficient $\gamma(t)$, the procedure of iteratively commuting the resulting new operators with all other 
previously computed will not close to a finite algebra (except, perhaps, for very special cases). In this case the resulting
extended algebra will probably contain an infinite number of degrees of freedom, a situation that presumably fits better with
the concept of thermal bath. A deeper study of this general case would be extremely interesting, at least in a perturbative way.

\appendix
\section{Appendix: Infinite-dimensional symmetry in the ``free'' damped particle}
\label{infinitegroup}

In this appendix, we turn our attention to the ``free'' damped particle as the
simplest case of physical system subjected to a dissipative force and perform a
similar analysis to that carried out in Subsection \ref{timesymmetry} for the
damped harmonic oscillator, forcing the introduction of the time generator in the symmetry algebra.

The basic operators for the damped particle (obtained as the $\omega\rightarrow 0$ of the damped harmonic oscillator,
see eq. (\ref{eq:P_DHO}-\ref{eq:X_DHO}))
\begin{equation}
 \hat P = -i \hbar \frac{\partial}{\partial x},\quad
 \hat X = x+\frac{i \hbar}{m \gamma}(1-e^{-\gamma t})\frac{\partial}{\partial x}
 \,.
 \label{eq:operatas DP}
\end{equation}

Renaming $\hat P\equiv\hat
P_0$, we introduce operators $\hat P_n$ and $\hat Y_n$ ($n$ being a non-negative integer)
\begin{align}
 \hat H_G &= i \hbar e^{\gamma t}\frac{\partial}{\partial t} & \hat H_{DP} &= i
\hbar \frac{\partial}{\partial t}\nonumber\\
 \hat P_n &= -i \hbar e^{- \gamma n t}\frac{\partial}{\partial x} & \hat Y_n
&=
i e^{- \gamma n t}\nonumber\\
\hat X &= x + \frac{i \hbar}{m \gamma}(1-e^{\gamma t})\frac{\partial}{\partial
x}\,.
\end{align}

It is interesting that they close an infinite-dimensional Lie algebra with non-null commutators:
\begin{align}
[\hat H_G,\hat P_n]& =-i \hbar \gamma n \hat P_{n-1}&
[\hat H_{DP},\hat P_n]& =-i \hbar \gamma n \hat P_{n} \nonumber \\
[\hat H_G,\hat X ]& =-i \frac{\hbar}{m}\hat P_0&
[\hat H_{DP},\hat X ]& =-i \frac{\hbar}{m}\hat P_1 \nonumber \\
[\hat X,\hat P_n]& = \hbar \hat Y_n& [\hat H_{DP},\hat Y_n]& =-i \hbar \gamma n \hat Y_{n} \nonumber\\
[\hat H_G,\hat Y_n]& =-i \hbar \gamma n \hat Y_{n-1}&  [\hat H_{DP},\hat H_G]& =i \hbar \gamma \hat H_G& \,, \label{gauge}
 \end{align}
containing a (centrally extended) Galilei algebra as
a subalgebra, for which $\hat Y_0 = i$ is the central generator. Even, the generators in the left column of (\ref{gauge}), for
$n=0,1$, also close a finite-dimensional
subalgebra in which $\hat H_G$ is a dynamical generator dual to $\hat Y_1$. 

It is interesting to look at the algebra (\ref{gauge}) as some sort of gauge algebra. In fact, we may start from the specific
realization $\hat H_G, \hat X, \hat P_0, \hat Y_0$  of the centrally extended Galilei algebra and try to turn it into 
a gauge (local) algebra by allowing the multiplication by certain functions of time (positive integer powers of $e^{-\gamma t}$) of the
space and time generators, and accordingly of the central one. This can be seen as the simplest symmetry of the free particle which
results as a consequence of requiring the gauge invariance under the generator  $e^{-\gamma t}\hat H_G$.

It would be worth investigating the possibility of repeating this brief analysis for the case of the symmetry of the harmonic oscillator.

\section*{Acknowledgments}

Work partially supported by the Fundaci\'on S\'eneca, Spanish MICINN and
Junta de Andaluc\'\i a under projects 08814/PI/08, FIS2011-29813-C02-01
and FQM219-FQM1951, respectively.

The authors wish to thank M. Calixto and D. Schuch for useful discussions and comments.

%

\end{document}